\documentclass[a4paper,fleqn,usenatbib]{mnras}

\usepackage{graphicx}	
\usepackage{amsmath}	
\usepackage{amssymb}	

\title[A low-mass triple system with a wide brown dwarf component]{A low-mass triple system 
       with a wide L/T transition brown dwarf component: NLTT~51469AB/SDSS~2131$-$0119}

\author[B. Gauza et al.]{
B. Gauza,$^{1,2}$\thanks{E-mail: bgauza@das.uchile.cl}\thanks{Based on observations made with the Gran Telescopio Canarias (GTC), installed 
   in the Spanish Observatorio del Roque de los Muchachos of the Instituto de Astrof\'isica de Canarias, in the island of
   La Palma (program GTC27-13B).}
V.~J.~S. B\'ejar,$^{2,3}$ 
A. P\'erez-Garrido,$^{4}$
N. Lodieu,$^{2,3}$
R. Rebolo,$^{2,3,5}$\newauthor
~M.~R. Zapatero Osorio,$^{6}$
B. Pantoja,$^{1}$
S. Velasco$^{2,3}$ 
and J.~S. Jenkins$^{1}$
\\
$^{1}$Departamento de Astronom\'ia, Universidad de Chile, Camino el Observatorio 1515, Las Condes, Santiago, Chile, Casilla 36-D\\
$^{2}$Instituto de Astrof\'isica de Canarias (IAC), Calle V\'ia L\'actea s/n, E-38200 La Laguna, Tenerife, Spain\\
$^{3}$Departamento de Astrof\'isica, Universidad de La Laguna (ULL), E-38206 La Laguna, Tenerife, Spain\\
$^{4}$Universidad Polit\'ecnica de Cartagena, Campus Muralla del Mar, Cartagena, Murcia E-30202 Spain\\
$^{5}$Consejo Superior de Investigaciones Cient\'ificas, CSIC, Spain\\
$^{6}$Centro de Astrobiolog\'ia (CSIC-INTA), Ctra. Ajalvir km 4, 28850 Torrej\'on de Ardoz, Madrid, Spain
}

\date{Accepted 2019 April 30. Received 2019 March 21; in original form 2018 October 4}

\pubyear{2019}

\begin{document}
\label{firstpage}
\pagerange{\pageref{firstpage}--\pageref{lastpage}}
\maketitle

\begin{abstract}
   We demonstrate that the previously identified L/T transition brown dwarf SDSS J213154.43$-$011939.3 (SDSS 2131$-$0119) 
   is a widely separated (82\farcs3, $\sim$3830\,au) common proper motion companion to the low-mass star NLTT~51469, which 
   we reveal to be a close binary itself, separated by 0\farcs64\,$\pm$\,0\farcs01 ($\sim$30\,au). We find the proper motion of 
   SDSS 2131$-$0119 of $\mu_{\alpha}\cos\delta$\,=\,$-$100\,$\pm$\,20~mas/yr and $\mu_{\delta}$\,=\,$-$230\,$\pm$\,20~mas/yr 
   consistent with the proper motion of the primary provided by {\it Gaia} DR2: 
   $\mu_{\alpha}\cos\delta$\,=\,$-$95.49\,$\pm$\,0.96 mas/yr and $\mu_{\delta}$\,=\,$-$239.38\,$\pm$\,0.96 mas/yr. Based 
   on optical and near-infrared spectroscopy we classify the primary NLTT 51469A as a M3\,$\pm$\,1 dwarf, estimate photometrically
   the spectral type of its close companion NLTT 51469B at $\sim$M6 and confirm the spectral type of the brown dwarf to be 
   L9\,$\pm$\,1. Using radial velocity, proper motion and parallax we derived the {\it UVW} Galactic space velocities of 
   NLTT 51469A, showing that the system does not belong to any known young stellar moving group. The high $V$, $W$ 
   velocities, lack of a 670.8\,nm Li\,{\sc i} absorption line, and absence of H{$\alpha$} emission, detected X-rays or UV excess, 
   indicate that the system is likely a member of the thin disk population and is older than 1 Gyr. For the parallactic 
   distance of 46.6$\pm$1.6~pc from {\it Gaia} DR2 we determined luminosities of $-1.50^{+0.02}_{-0.04}$ and $-4.4\pm0.1$~dex 
   of the M3 and L9, respectively. 
   Considering the spectrophotometric estimation which yields a slightly lower distance of $34^{+10}_{-13}$~pc the obtained luminosities
   are $-1.78^{+0.02}_{-0.04}$ and $-4.7^{+0.3}_{-0.5}$~dex.
   We also estimated their effective temperatures and masses, and obtained 3410$^{+140}_{-210}$~K 
   and 0.42\,$\pm$\,0.02~$M_{\odot}$ for the primary, and 1400--1650\,K and 0.05--0.07~$M_{\odot}$ for the wide companion. 
   For the $\sim$M6 component we estimated $T_{\rm eff}$\,=\,2850\,$\pm$\,200~K and $m$\,=\,0.10$^{+0.06}_{-0.01}$~$M_{\odot}$.
\end{abstract}




\section{Introduction}

Brown dwarfs are objects that have insufficient mass to sustain stable nuclear fusion in their interiors. After they are 
formed they evolve getting fainter and cooler. The emergent spectra of brown dwarfs were found so different from that of 
the latest type M dwarf stars that the establishment of two new spectral types, L and T \citep{1999ApJ...519..802K, 
2006ApJ...637.1067B}, was indispensable to properly classify them. The L-type class encompasses the least massive stars and 
most massive brown dwarfs and spans effective temperatures ($T_{\rm eff}$s) of approximately 2500 to 1300\,K
\citep[e.g.,][]{1999ApJ...519..802K, 2005ARA&A..43..195K}. The T-type class includes solely brown dwarfs with temperatures 
below $\sim$1400\,K and down to 500--700\,K \citep[e.g.,][]{2004AJ....127.3516G, 2006ApJ...639.1095B, 2006ApJ...637.1067B, 
2008MNRAS.391..320B, 2009ApJ...695.1517L}. Most recently, searches using the space based Wide-field Infrared Survey Explorer 
(WISE; \citealt{2010AJ....140.1868W}) data revealed a population of even lower $T_{\rm eff}$ objects from approximately 
500--700\,K down to about 250\,K, which have been classified as Y dwarfs \citep[e.g.][]{2011ApJ...743...50C, 
2012ApJ...753..156K, 2014ApJ...786L..18L, 2017ApJ...842..118L}.

The L to T-type transition undergoes a dramatic change in near-infrared (near-IR) $J-K_s$ colour, which becomes 
bluer by about two magnitudes as part of a major change in spectral morphology. This occurs over a narrow range in 
temperature $\Delta T_{\rm eff}\sim$200--300\,K \citep[e.g.,][]{2004AJ....127.3516G}. This dynamic phenomenon is still 
not entirely reproduced and matched by theoretical models, nevertheless the general description of the L/T transition is 
thought to be fairly well understood for high-gravity atmospheres. It is physically characterised by the break up and 
clearing of the condensates clouds and a sudden sedimentation of condensed species below the photosphere when the local 
temperature is $\lesssim$1,500\,K, and by the appearance of methane absorption in the 1.0--2.5\,$\mu$m region 
\citep{1997ARAA..35..137A, 2001ApJ...556..357A, 2000ApJ...542..464C, 2001ApJ...556..872A, 2006ApJ...640.1063B, 
2008ApJ...689.1327S}.

Since the discoveries of the first unambiguous brown dwarfs in the mid 1990s \citep{1995Natur.377..129R, 1995Natur.378..463N} 
large area imaging surveys in optical and near-infrared wavelengths like, among others, DENIS \citep{1994ApSS.217....3E}, 
2MASS \citep{2006AJ....131.1163S}, SDSS \citep{2000AJ....120.1579Y, 2003AJ....126.2081A} and more recently UKIDSS 
\citep{2007MNRAS.379.1599L}, {\it WISE} \citep{2010AJ....140.1868W} and Pan-STARRS \citep{2002SPIE.4836..154K} enabled 
significant growth in the population of known brown dwarfs with the current number being around 2000 objects 
\citep{1997AA...327L..25D, 2005ARA&A..43..195K, 2011ApJS..197...19K, 2012AA...541A.163S, 2014ApJ...792..119D, 
2016ApJ...830..144R, 2017MNRAS.469..401S}. However, a great majority of these are single objects and only a small 
fraction are found to be components in binary or multiple systems \citep[e.g.,][]{2010AJ....139..176F, 2014ApJ...792..119D,
2015ApJ...802...37B, 2016AA...587A..51S}.

Ultracool companions to stars are of interest because both components share the same age, metallicity and distance, 
which are easier to determine for the brighter primary. Substellar objects with well constrained age and metallicity are 
valuable reference points for calibrating evolutionary and atmospheric models \citep[e.g.,][]{2006MNRAS.368.1281P, 
2010AJ....139..176F, 2012MNRAS.420.3587J, 2013MNRAS.431.2745G, 2017MNRAS.470.4885M}. Widely separated substellar 
companions are particularly useful for characterisation because they can be directly observed using seeing-limited 
instruments.

In this paper we present a nearby, common proper motion triple system identified in our search for ultracool companions 
to stars using the VISTA Hemisphere Survey data. We describe the search and identification methods in Section 2. Section 3 
contains the description of follow-up observations aimed at confirming the companionship and characterisation of the objects. 
Next, we determine the spectral types of the components in Section 4 and demonstrate the binarity of the NLTT 51469 star 
in Section 5. In Section~6 we compare the distance measured by {\it Gaia} to the spectrophotometric distance obtained for 
the primary and the L9 companion. Section 7 contains a discussion on the probability of chance alignment of the two wide 
components. In Section 8 we determine the physical properties of the objects, including radial velocity, galactic kinematics 
and constraint on the age of the system, luminosities, masses and temperatures of the individual components. Conclusions 
are future prospects are presented in Section 9.

\begin{figure}
 \centering
 \includegraphics[scale=0.65,keepaspectratio=true]{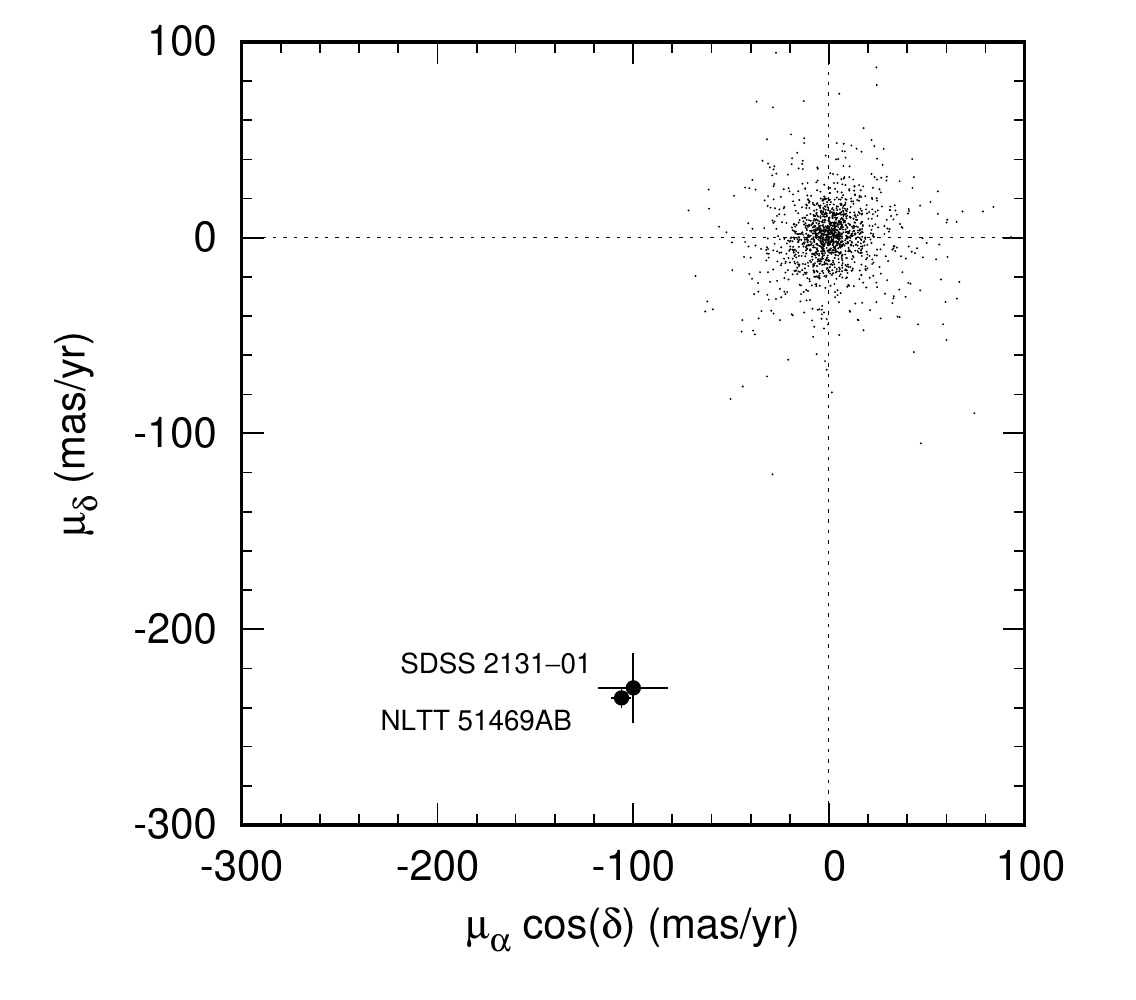}
 \caption{Proper motion vector-point diagram for NLTT~51469 and its companion.
 All correlated objects within 20 arcmin from the primary with $J$\,$<$\,17.5~mag are 
 plotted as black dots, while the two components of the common proper motion pair are 
 plotted with larger points and labelled. Error bars correspond to the astrometric rms 
 of stars with similar brightness as of the components.}
 \label{fig_pm}
\end{figure}

%

\section{Search and identification}
The VISTA Hemisphere Survey (VHS; \citealt{2004Msngr.117...27E}) is an ongoing imaging survey for which one of the main 
goals is the detection of very low-mass stars and substellar objects. The VHS will map the entire southern hemisphere of the sky 
($\sim$20,000\,deg$^2$), with the exception of the areas within the VISTA Kilo-Degree Infrared Galaxy Survey (VIKING) and the 
Variables in the Via Lactea (VVV) survey. Observations covering an area of 10,000\,deg$^2$ are carried out in two near-IR 
bands, $J$ and $K_s$, 5000\,deg$^2$ in $JHK_s$ and another 5000\,deg$^2$ in $YJHK_s$, reaching a median 5$\sigma$ point source 
detection limits of $J$\,=\,20.2 and $K_s$\,=\,18.1~mag \citep{2013Msngr.154...35M}. The VHS catalog provides astrometry and 
photometry in the $J$, $K_s$ bands, and $Y$, $H$ bands when available. The VISTA photometric system is calibrated using the 
magnitudes of color-selected 2MASS stars converted onto the VISTA system using color equations, including terms to account for 
interstellar reddening\footnote{\url{http://casu.ast.cam.ac.uk/surveys-projects/vista/technical/photometric-properties}}. 
Photometric calibrations are determined to an accuracy of 1-2\%. The astrometric solution for VHS observations is computed by 
the automatic pipeline of the survey, using point sources from the 2MASS catalog. The world coordinate system of VISTA images 
is calibrated to 0.1--0.2~arcsec accuracy.

As of the early data releases from VHS prior to March 2013 which covered about 8500~deg$^2$ we built a catalog of high proper 
motion objects in the southern hemisphere, with motions greater than 0.15--0.20 arcsec per year. We combined VHS catalog 
measurements (astrometry and {\it YJHK} photometry) with 2MASS point source catalogue measurements (as a reference epoch) to 
identify moving objects in the sky areas overlapping with VHS. The time baseline between the two surveys ($>$10 years) allow 
to measure proper motions with precision of 10\,mas/yr or better. Our search was restricted to sources with 
$J_{\rm 2MASS}$\,$\leq$\,17.5~mag, and fainter than $J_{\rm VHS}$\,=\,11~mag to 
exclude objects affected by saturation from the VHS database. We also cross-matched VHS sources with WISE All-Sky and USNO-B1.0 
catalogs to get mid-infrared and optical photometry information and to filter out contaminants based on mid-infrared and 
optical constraints where available.

In the cross correlation of VHS and 2MASS catalogs we have found 50382 objects with $J$\,=\,11--17.5~mag range and proper motions 
of $\mu\gtrsim$\,150~mas/yr taking into account the baseline of 10--12 years between these two surveys. Most of these high proper 
motion objects are relatively nearby M dwarfs with estimated photometric distances within 100~pc. The general catalogue is 
currently being developed and prepared for publication (A. P\'erez-Garrido et al. 2019, in preparation). Among this list, we have 
searched for objects co-moving with stars from the revised version of the New Luyten Catalogue of Stars With Proper Motions Larger 
than Two Tenths of an Arcsecond (NLTT; \citealt{2003ApJ...582.1001G}; \citealt{2003ApJ...582.1011S}). We required proper motions 
to be consistent within 50 mas/yr in both right ascension and declination to identify proper motion pairs. This corresponds to a 
factor of 2.5--3 times larger than the quoted 1\,$\sigma$ astrometric error of the surveys, which improves the scope for 
identifying moving objects fainter than the 2MASS completeness limits. The cross-match was limited to angular separations of 
20 arcmin radius.

\begin{figure*}
 \centering
 \includegraphics[scale=0.7,keepaspectratio=true]{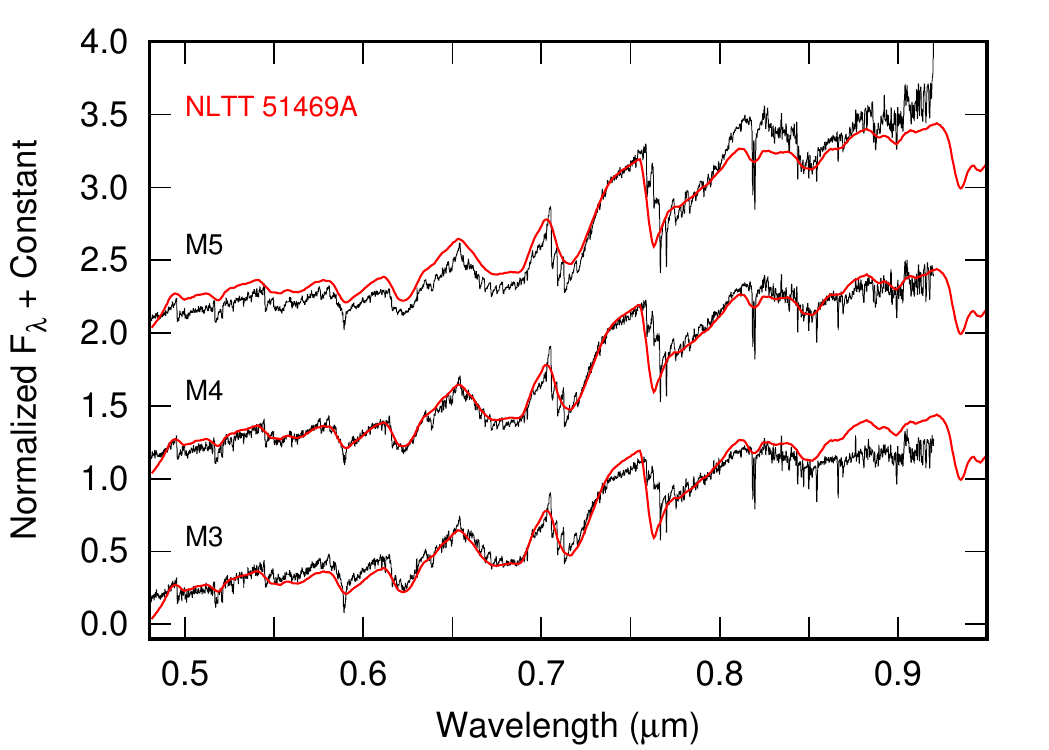}
 \includegraphics[scale=0.7,keepaspectratio=true]{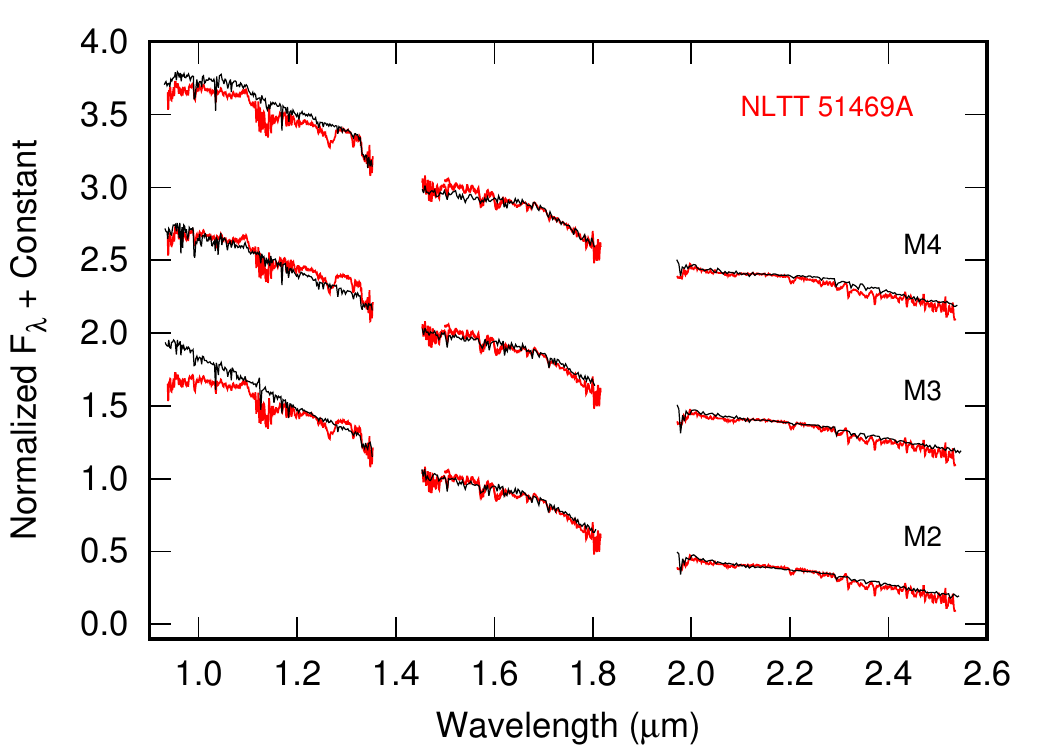}
 \caption{Low-resolution GTC/OSIRIS optical (left panel) and NTT/SofI near-infrared (right panel) spectra 
 of the primary NLTT 51469A plotted with red lines, compared to a grid of M2-M5 spectral templates with 
 labels indicating their type. The sources and references of used templates are described in Section \ref{spectypes}. 
 Spectra were normalized at 0.9\,$\mu$m at optical and at 1.65\,$\mu$m at near-IR and offset by a constant for display.}
 \label{fig_spectra_primary}
\end{figure*}

From the more than a hundred potential binaries and multiples we build a sample of candidate systems containing 
one or more components with near- and mid-IR photometric colors consistent with mid-M and later spectral types 
($J-K_s$\,$>$\,0.8, $J-H$\,$>$\,0.5, $J-W$2\,$>$\,1.5, $W$1$-W$2\,$>$\,0.5). Next, we verified each selected system, 
by checking for the consistency in distance modulus of its candidate components, estimated roughly from the apparent 
$J$-band brightness and the expected spectral type based on photometric colors. The color-spectral type and color-absolute 
magnitude relations were adopted from \cite{2006AJ....131.2722C, 2011ApJS..197...19K} and \cite{2012ApJS..201...19D}.

One of the identified pairs was the star NLTT~51469 and the brown dwarf SDSS J213154.43-011939.3 with on-sky 
separation of 82.3 arcsec. In the VHS-2MASS cross-match, we obtained proper motions of $\mu_{\alpha}\cos\delta$, 
$\mu_{\delta}$ = $-$82\,$\pm$\,16, $-$229\,$\pm$\,18 and $-$60\,$\pm$\,19, $-$270\,$\pm$\,20~mas\,yr$^{-1}$, for 
the primary and companion, respectively. The values are consistent within 50 mas\,yr$^{-1}$ as required, but the 
differences are around 1$\sigma$ error. We note that the centroid positions are likely unreliable, because of the 
brightness range of the surveys. On the one hand, the primary is out of the linear range of the VHS and on the other 
hand, the secondary is close to detection limit in 2MASS. Therefore, we employed also the SDSS astrometry, and 
obtained more precise proper motions by combining 2MASS and SDSS measurements for the M3 star, and VHS and SDSS 
measurements for the L9 brown dwarf. The resulting values are $\mu_{\alpha}\cos\delta$, $\mu_{\delta}$ =
$-$106\,$\pm$\,8, $-$235\,$\pm$\,8~mas\,yr$^{-1}$ for the M3 and $-$100\,$\pm$\,15, $-$230\,$\pm$\,15~mas\,yr$^{-1}$ 
for the L9. This is also in good agreement with the {\it Gaia} Data Release~2 measurement for the primary, which 
gives $\mu_{\alpha}\cos\delta$, $\mu_{\delta}$ = $-$95.49\,$\pm$\,0.96, $-$239.38\,$\pm$\,0.96~mas\,yr$^{-1}$ 
\citep{2016AA...595A...1G, 2018AA...616A...1G}. To within the quoted uncertainties both objects share the same 
proper motion, which is shown in Figure~\ref{fig_pm}. These proper motions differ significantly ($>$10$\sigma$) from 
the population of background field stars (with $J$\,$<$\,17.5~mag and within 20 arcminutes) also shown in the figure.

\section{Follow-up observations and data reduction}

\subsection{Near-infrared spectroscopy}
%
We obtained low-resolution near-IR spectroscopy of the primary NLTT~51469A using the Son of ISAAC (SofI) 
spectro-imager installed on the NTT on 16 November 2013 (programme ID: 092.C-0874(B), PI: Gauza). SofI is 
equipped with a Hawaii HgCdTe 1024$\times$1024 array with 18.5\,$\mu$m pixels. We used the large-field mode, 
offering a field-of-view of 4.9$\times$4.9 arcmin with a 0.288 arcsec pixel scale, and blue (950--1640 nm) and 
red (1530--2520 nm) grisms combined with a slit of 1 arcsec oriented to the parallactic angle. This configuration 
provides a resolving power of $R$\,$\sim$\,550. We used single integrations of 60\,s and 120\,s for blue and 
red grism, respectively, repeated in an ABBA pattern for both configurations to remove the sky contribution. To 
correct for telluric absorption features we observed an early type hot star with the same configuration (HD~13936; 
$J$\,=\,6.46 mag; A0V; \citealt{2007AA...474..653V}) shortly after NLTT~51469A but at a lower airmass (1.2 vs. 2.4). 
The sky conditions during the observations were clear with a seeing around 1 arcsecond. During the afternoon 
preceding the observations we acquired standard calibration frames for data reduction, including bias, spectral 
flats and Xenon arcs.

We used the ESO SofI pipeline recipes version 1.5.5 within the {\it Gasgano} tool to reduce the raw data and to 
align and combine the dispersed images from the four ABBA positions along the slit to obtain images of the 2D spectra.
We then extracted the spectra using standard routines under the {\sc apall} task in {\sc iraf} and 
wavelength calibrated it via Xenon arc lines. The dispersion solution had an rms of 0.54 and 0.62\,\AA ~for the 
blue and red part of the spectrum, respectively, and the resolution of the spectra was 24\,\AA~($R$\,$\sim$\,530) and 
35\,\AA~($R$\,$\sim$\,580) in the blue and red arm, respectively. We corrected for telluric lines, dividing the spectra 
by the A0V standard HD~13936 and multiplying by a blackbody of a corresponding effective temperature of 9700~K.

\subsection{Optical spectroscopy}
%
We performed long-slit, low resolution optical spectroscopy of NLTT 51469 and its wide companion using the OSIRIS
instrument \citep[Optical System for Imaging and low-intermediate Resolution Integrated Spectroscopy;][]{2010hsa5.conf...15C} 
at the Gran Telescopio de Canarias (GTC) telescope located on the Observatorio del Roque de los Muchachos (island of La Palma, 
Spain). OSIRIS is equipped with two 2048$\times$4096 Marconi CCD42-82 detectors which provides a field-of-view approximately 
7\,$\times$\,7~arcmin$^2$ with an unbinned pixel scale of 0.125~arcsec. 

\begin{figure}
 \centering
 \includegraphics[scale=0.95,keepaspectratio=true]{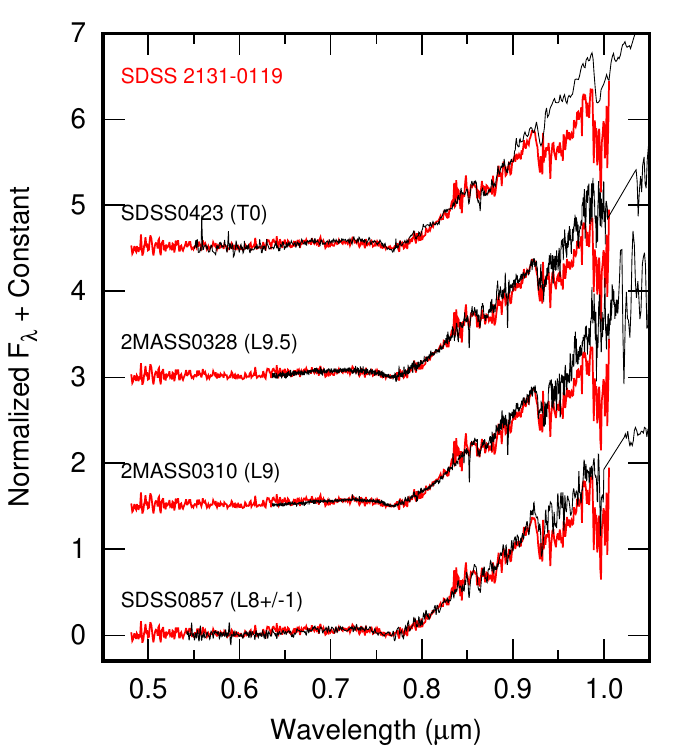}
 \caption{GTC/OSIRIS low-resolution optical spectrum of the brown dwarf companion SDSS~2131--0119 (red lines), overplotted with a set 
 of spectra of known L8-T0 field dwarfs (black lines). Names and spectral types of used templates are labelled in the plot and 
 corresponding references are given in Section~\ref{spectypes}. All spectra are normalized at 0.9\,$\mu$m and shifted by a constant.}
 \label{fig_spectra_comp}
\end{figure}

We observed both objects using a R300R grating with a slit width of 2.5 arcsec and 2\,$\times$\,2 binning, which allowed 
us to measure the general spectral energy distribution at a resolution of 87\AA~($R$\,$\sim$\,76), covering the 0.5--1.0\,$\mu$m 
range. Observations were acquired on October 26, 2013 as part of the GTC27-13B program (PI: N. Lodieu). Single exposures of 
15 and 600\,s were obtained for the primary and secondary, respectively. Bias frames, continuum lamp flat fields, and Xenon, 
Neon and Argon arcs were obtained during the afternoon preceding the observations. The spectrophotometric standard star 
G158-100 \citep{1984PASP...96..530F, 1990AJ.....99.1621O} was observed on the same night as the scientific target, first using 
the same spectroscopic setup as for the target, and then also with a broad z-band filter to correct for second-order contamination 
beyond 9200\,\AA~(see procedure in \citealt{2014AA...568A...6Z}).

The OSIRIS data were reduced with standard procedures using routines within {\sc iraf}. The raw spectra were bias-corrected, 
trimmed and divided by a normalized continuum lamp flat field. From the 2D images we extracted the spectra using the {\sc apall} 
routine and calibrated in wavelength with the lines from combined XeNeAr arc lamps. The correction for instrumental response 
was applied using a response function generated from the spectrophotometric standard star.

To look for spectral signatures of age like e.g., the lithium Li\,{\sc i} line at 670.8\,nm or H$\alpha$ emission at 656.3\,nm
and also to obtain additional measurement of radial velocity of the primary star, we acquired intermediate-resolution 
($R$\,$\sim$\,5000) optical spectroscopy of NLTT~51469 using the Fibre-fed RObotic Dual-beam Optical Spectrograph (FRODOSpec; 
\citealt{2004AN....325..215M}) on the robotic Liverpool Telescope (LT; \citealt{2004SPIE.5489..679S}) in La Palma. The FRODOSpec 
is a 12\,$\times$\,12 fibre integral-field unit spectrograph for the LT, designed mainly to study point sources. Each fibre covers 
a field of view on sky of $\sim$0.83\arcsec\,$\times$\,0.83\arcsec, corresponding to a total field of view of approximately 
10\arcsec\,$\times$\,10\arcsec\,\citep{2004AN....325..215M, 2012AN....333..101B}. It uses a dichroic beam-splitter to separate the 
incident light at around 5750\AA, into the blue and red arm, covering in the high resolution mode 3900--5100\AA~and 5900-8000\AA, 
respectively.

We used the Volume Phase Holographic (VPH) grating available on the instrument, which provides higher resolution ($R$\,$\sim$\,5300 in 
red arm) than the conventional diffraction grating. Observations of NLTT~51469 were performed on July 24, 2016. Two individual 
exposures of 600\,s integrations were collected. A radial velocity standard star GJ~873 was observed with the same instrumental 
setup and 30\,s integrations on 20 August, 2016. The Xenon arcs and Tungsten lamp exposures were acquired prior to each target.
Raw FRODOSpec data were reduced by two sequentially invoked automatic pipelines. The first one, known as the L1, processes the 
CCD images performing bias subtraction, overscan trimming and flat fielding. The second one (L2) performs the processing 
appropriate to integral field spectra reduction. Details of the processing steps of the two pipelines are described in 
\cite{2012AN....333..101B}. The reduced data products contain a snapshot of the data taken at key stages in the reduction 
process, including the final sky subtracted and wavelength calibrated 1D spectra. No correction of the instrumental response
was applied.

Since the KI line at 766.5~nm is strongly affected by telluric lines we applied a telluric correction to measure the pEW of these 
lines. For this purpose, we divided the spectrum of the target by the spectrum of a hot, early type star (BD+28 4211, sdO) 
observed on the same night. The spectrum of telluric star was first normalized in the whole spectral range except parts affected 
by telluric absorption. We only used this telluric corrected spectrum for this purpose, because the S/N of the telluric star
spectrum was poor and introduced more noise to the rest of the spectrum of our target.

\subsection{NOT/FastCam lucky imaging}
On November 16th 2016, we collected 5,000 individual frames of NLTT~51469 in the {\it I} band using the lucky imaging FastCam 
instrument \citep{2008SPIE.7014E..47O} at the 2.4m Nordic Optical Telescope (NOT) at the Observatorio del Roque de Los Muchcachos 
in La Palma, with 30 ms exposure time for each frame. FastCam is an optical imager with a low noise EMCCD camera which allows to 
obtain speckle-featuring not saturated images at a high frame rate. In order to construct a high spatial resolution, diffraction limited, 
long-exposure image, the individual frames were bias subtracted, aligned, and co-added using our own lucky imaging algorithm 
\citep{2011AA...526A.144L, 2016MNRAS.460.3519V}. Figure~\ref{figfc} presents the high resolution image constructed by co-addition 
of the best percentage of the images using the lucky imaging and shift-and-add method and processed with the wavelet filtering 
algorithm. Owing to the atmospheric conditions of the night, the selection of 10\% of the individual frames was found to be the best 
solution to produce a deep and diffraction limited image of the target, resulting in a total integration time of 15~s. To calibrate
the plate scale and orientation we used observations of the M15 globular cluster core performed on the same night compared against 
the {\it HST} WFPC2 catalog of M15 \citep{2002AJ....124.3255V}.

\subsection{Clay/MagAO imaging}
We observed NLTT 51469 using the Magellan Clay telescope at Las Campanas Observatory in Chile on the night of May 2, 2018. 
Observations were performed in the $J$-band using the infrared camera Clio-2 in the narrow mode which provides a
field-of-view of 16\arcsec\,$\times$\,8\arcsec \citep{2015ApJ...815..108M}. We nodded in an ABBA 
pattern to subtract the background and obtained 5 frames of 10\,s integration at each nod position. In the same manner we 
acquired shallow, unsaturated images with shorter, 0.5\,s individual integrations for the photometry. Shortly after the 
science target we observed the binary star 70~Oph which has an accurately determined orbit \citep{2000A&AS..145..215P} for 
the calibration of the pixel scale and orientation.

\begin{figure}
 \centering
 \includegraphics[scale=0.8,keepaspectratio=true]{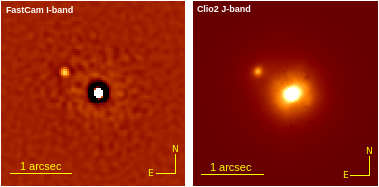}
 \caption{Resolved images of the central binary NLTT~51469AB. {\it Left:} The NOT/FastCam $I$-band image after 
 lucky imaging processing and the wavelet filtering. {\it Right:} The MagAO/Clio2 $J$-band image. The companion 
 is detected at 0.63\,arcsec separation and position angle of $\sim$57\,deg. The displayed field of view is 
 3\arcsec\,$\times$\,3\arcsec.}
 \label{figfc}
\end{figure}

We reduced the raw data cubes using {\sc PyRAF} routines, which included subtraction of the proper nod-pairs to remove 
the sky contribution, registering, aligning and median stacking of the individual frames. From the obtained image of 70~Oph 
we calculated the plate scale of 15.523\,$\pm$\,0.254 mas/pix and the NORTHClio angle of $-$1.78\,$\pm$\,0.40~deg. The 
NORTHClio angle is then used to find the derotation angle needed to get the North-up and East-left orientation. These 
values are consistent with the nominal values of 15.846\,$\pm$\,0.064 mas/pix and NORTHClio\,=\,$-$1.797\,$\pm$\,0.34\,deg 
provided in \cite{2015ApJ...815..108M}. A 3\,$\times$\,3 arcsec cut-off of the final derotated image of the pair 
is presented in Figure~\ref{figfc}

\subsection{VLT/NACO imaging}
We also acquired near-IR $J, H, K_s$ images of the primary using the NACO instrument, short for the Nasmyth Adaptive 
Optics System (NAOS, \cite{2003SPIE.4839..140R}) of the VLT-UT1, coupled to the CONICA high contrast infrared camera 
\citep{1998SPIE.3354..606L}. Observations were completed on October 24, 2018 (programme ID 0102.C-0899(A), PI Gauza), 
at average airmass of 1.1 and with thin cirrus clouds. We used the infrared wavefront sensor with the N90C10 dichroic, 
and our target star as a natural guide star. We chose high sensitivity mode of CONICA with the FowlerNsamp read-out 
and the S13 objective which provides 14\arcsec$\times$14\arcsec~field of view and the smallest available pixel scale 
of 13.26\,$\pm$\,0.03 mas/pix \citep{2003A&A...411..157M}. Individual exposures of 60 and 20\,s were taken in a 5 and 
9 position jittering pattern for the $J$ and $H$, $K_s$ filters, respectively, for a proper sky background subtraction.
In the same observing block, with the same setup and observing technique we also obtained $K_s$ band images of a 
known binary star WDS\,20204+0118 \citep{2001AJ....122.3466M} for pixel scale and orientation calibration.

Raw images were processed using the ESO NACO pipeline kit version 4.4.6, run within the Gasgano software tool, 
version 2.4.8. This included the dark and flat field corrections, sky subtraction, alignment of individual frames 
and stacking of each frameset. Astrometric and photometric measurements obtained from the final reduced $JHK_s$ 
images are described in Section~5.

\section{Spectral classification}
\label{spectypes}
We based the spectral type determination for the two components of this system on the low-resolution 
spectra, optical and near-IR for the primary, and optical for the wide companion. To classify the objects in a qualitative manner, 
we went through a direct visual comparison of the spectra with a set of known field dwarf spectral templates, separately in the 
optical and near-IR regimes. The optical spectra of known M dwarf templates were retrieved from the Sloan Digital Sky Survey 
(SDSS; \citealt{2000AJ....120.1579Y}) spectroscopic database provided by \cite{2007AJ....133..531B}. This database contains 
a repository of good-quality composite spectra of low-mass dwarfs (M0--L0), one per subclass, spanning the 380--940\,nm 
wavelength range. In the near-IR we used the M-dwarf spectral templates available in the IRTF Spectral Library\footnote{\url
{http://irtfweb.ifa.hawaii.edu/~spex/IRTF_Spectral_Library/}}, which provides $R$\,$\sim$\,2000 spectra with S/N\,$\gtrsim$\,100 
over the 0.8--2.5\,$\mu$m range.

In Figure~\ref{fig_spectra_primary} we display the optical and near-IR spectra (left and right panel, respectively) of NLTT~51469A
overplotted with a grid of best matching spectral templates. Regions contaminated by strong telluric absorption in the near-IR 
spectra around 1.4 and 1.9~$\mu$m were not considered in the comparison and are cleared for display. Also, the region at
$\sim$1.15~$\mu$m (in the $J$-band) appears to be affected by poor telluric correction. This may be due to the difference in airmass 
between the target and the standard star, nonetheless this does not preclude the spectral type classification which makes use of the 
full spectral range. The M dwarfs used as comparison objects in the near-IR are HD~95735 (M2V), Gl~388 (M3V) and Gl~213 (M4V) 
\citep{2009ApJS..185..289R, 2005ApJ...623.1115C}. Figure~\ref{fig_spectra_comp} contains a comparison of optical spectrum of the 
brown dwarf co-moving with NLTT~51469A with a grid of L8--T0 field dwarf spectra. The known objects used as templates are 
SDSS J085758.45+570851.4 (L8\,$\pm$\,1), SDSS J083008.12+482847.4 (L9\,$\pm$\,1), 2MASS J03284265+2302051 (L9.5\,$\pm$\,0.5) and 
SDSS J042348.57-041403.5AB (T0\,$\pm$\,0.5) and their data were taken from \cite{2004AJ....127.3516G, 2004AJ....127.3553K} 
and \cite{2006AJ....131.2722C}. The broad feature shortward of 1 micron in the spectrum of the object is due to lack of 
telluric correction.

For the primary we determine a spectral type of M3.0 dwarf with a one subclass uncertainty, considering both optical and near-IR 
spectra. For the companion we assign a spectral type of L9, with a one subclass uncertainty, using its optical spectrum. The M3 
primary does not show strong molecular absorption due to CaH in the optical spectrum, thus indicating that this star is not a 
metal-depleted source, which contrasts with the results of the RAVE 4th data release catalog (\cite{2013AJ....146..134K}; 
[m/H]\,=\,$-$2.31\,$\pm$\,0.18). Indeed, as shown in Figure~\ref{fig_spectra_primary}, the optical spectrum of the primary is nicely 
reproduced by field, high-gravity solar-metallicity stars. The L9\,$\pm$1 type is consistent with the previous determination by 
\cite{2006AJ....131.2722C} who classified the object using 0.8--2.5\,$\mu$m near-IR spectra.

The optical/near-IR/mid-IR multi-band colours of the primary: $V-I$\,=\,2.38\,$\pm$\,0.03, $I-J$\,=\,1.35\,$\pm$\,0.05, 
$J-K_{\rm s}$\,=\,0.80\,$\pm$\,0.05, $K_{\rm s}-W1$\,=\,0.15\,$\pm$\,0.05 mag are compatible with the colors of M2--M4V spectral 
type standards \citep{1994AJ....107..333K, 2013ApJS..208....9P}. The brown dwarf companion, with $z-J$\,=\,2.66\,$\pm$\,0.12,
$J-H$\,=\,0.81\,$\pm$\,0.03, $J-K_{\rm s}$\,=\,1.43\,$\pm$\,0.03 and $J-W2$\,=\,2.41\,$\pm$\,0.09~mag also shows an agreement 
with the typical colors of field age late-L dwarfs, however a large scatter of $\sim$0.5~mag and similarity of these colors 
of L5--T0 objects \citep{2006AJ....131.2722C, 2012ApJS..201...19D} prevents a more precise distinction of the spectral type 
from photometry.

\begin{figure}
 \centering
 \includegraphics[scale=0.8,keepaspectratio=true]{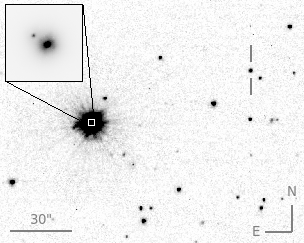}
 \caption{VISTA $J$-band image of the NLTT~51469/SDSS~2131$-$0119 system, with the location of the brown dwarf component 
 marked with two vertical lines.  The field shown is 3\arcmin\,$\times$\,2.5\arcmin~and oriented North up and East to the left.
 The zoomed-in region showing the central binary is the 3\arcsec\,$\times$\,3\arcsec~MagAO/Clio2 image.}
 \label{fig_fc}
\end{figure}

\begin{table*}
\centering
\caption{Photometry and astrometry of NLTT~51469AB \label{phot_AB}}
\begin{tabular}{l c c c c c}
\hline
\hline
 Instr., band & mag(A) & mag(B) & $\rho$ (\arcsec) & $\theta$ (deg) & Epoch (MJD)\\
\hline
FastCam $I$ & 11.25\,$\pm$\,0.30\hspace{1.5mm}  & 14.55\,$\pm$\,0.30   & 0.625\,$\pm$\,0.009 & 57.28\,$\pm$\,0.76 & 57708 \\
Clio-2 $J$  &  9.98\,$\pm$\,0.06  & 12.21\,$\pm$\,0.06   & 0.628\,$\pm$\,0.012 & 56.59\,$\pm$\,0.42 & 58240 \\
NACO $J$    & 9.975\,$\pm$\,0.031 & 12.245\,$\pm$\,0.031 & \hspace{1.5mm}0.641\,$\pm$\,0.003$^a$ & \hspace{1.5mm}55.95\,$\pm$\,0.16$^a$ & 58415 \\
NACO $H$    & 9.399\,$\pm$\,0.023 & 11.751\,$\pm$\,0.023 & \hspace{1.5mm}0.638\,$\pm$\,0.003$^a$ & \hspace{1.5mm}56.15\,$\pm$\,0.14$^a$ & 58415 \\
NACO $K_s$  & 9.163\,$\pm$\,0.025 & 11.513\,$\pm$\,0.025 & 0.638\,$\pm$\,0.003 & 55.92\,$\pm$\,0.13 & 58415 \\
\hline
\end{tabular}\\
$^a$ using pixel scale and orientation calibration from \cite{2016A&A...587A..35K}.
\end{table*}

\begin{table}
\centering
\caption{Spectrophotometric distance estimates for SDSS~2131--0119 \label{specphot_distance}}
\begin{tabular}{l l c c}
\hline
\hline
 Method & Ref. & d (pc) & $\Delta$/$\sigma$ \\
\hline
$M_J$ vs. SpT    & F15 & 38 (+7, -11)  & 0.94 \\
$M_H$ vs. SpT    & F15 & 39 (+7, -11)  & 0.83 \\
$M_{Ks}$ vs. SpT & F15 & 41 (+6, -12)  & 0.61 \\
$M_{W1}$ vs. SpT & F15 & 40 (+14, -11) & 0.52\\
$M_{W2}$ vs. SpT & F15 & 46 (+14, -11) & 0.05\\
$M_J$ vs. SpT    & D12 & 33 (+12, -8)  & 1.34\\ 
$M_{W2}$ vs. SpT & D12 & 44 (+15, -11) & 0.20\\
\hline
Mean    &  & 40.1\,$\pm$\,10.7 & 0.64\\
\hline
\end{tabular}
\begin{flushleft}
References. -- (F15): \cite{2015ApJ...810..158F}; (D12): \cite{2012ApJS..201...19D}
\end{flushleft}
\end{table}

\section{Binarity of the NLTT~51469 star}
We have analysed the FastCam, Clio-2 and NACO images at the three epochs spanning a 1.9~yr baseline. Using imcentroid and 
adopting the instrument angle and pixel scale of FastCam at NOT of 30.5\,$\pm$\,0.1~mas/pix determined from astrometric 
calibrations using the M15 cluster images, we have measured a relative angular separation between the NLTT~51469A star and 
the additional nearby source of $\rho$=0.625\,$\pm$\,0.009~arcsec and a position angle of $\theta$=57.28\,$\pm$\,0.76\,deg. 
From the Clio-2 observations, using pixel scale and orientation calibrations from \cite{2015ApJ...815..108M} we have 
measured $\rho$=0.6414\,$\pm$\,0.0001~arcsec and $\theta$=56.57\,$\pm$\,0.13~deg and using our calibrations with 70 Oph, 
$\rho$=0.628\,$\pm$\,0.012~arcsec and $\theta$=56.59\,$\pm$\,0.42~deg. From the NACO $K_s$-band image we have measured 
$\rho$=0.638\,$\pm$\,0.003~arcsec and $\theta$=55.92\,$\pm$\,0.13~deg with the instrument angle and 13.24\,$\pm$\,0.04~mas/pix 
pixel scale calibrated using the WDS\,20204+0118 image in the same band. Considering the proper motion of the primary and the 
time span from FastCam observation the $\rho$ and $\theta$ would be 1.07\,$\pm$\,0.02~arcsec and 41.6\,$\pm$\,0.8~deg if the
secondary was a non-related background object.

From these images, we have also measured the relative flux between the two objects using the peak value ratio and psf
photometry using the {\sc daophot} package and determined the relative magnitude difference in each of the four bands.
To account for the additional source, which is unresolved in 2MASS and DENIS, we deblended the catalog magnitudes to give 
the appropriate values of individual components. The magnitudes, angular separations and position angles are listed 
in Table~\ref{phot_AB}.

Considering the relative high proper motion of the primary star of 260 mas/yr, and the expected magnitude of the secondary 
star of $I$\,$\sim$14.5\,mag, the fainter component should have been detected at the same sky position in the old photographic 
plates of Digital Sky Survey images if it was a stationary background object. Having the three epoch images we prove 
that this close pair is indeed co-moving, as the measured angular separations and position angles remain consistent within 
1.5$\sigma$. Moreover, the $\rho$ and $\theta$ measured on the earliest and latest epoch observations differ by approximately 
40$\sigma$ and 20$\sigma$, respectively, from the values expected in case the secondary was a stationary object.

The probability that these two stars are found in such close proximity by chance is very low. Given the 0.6~arcsec separation 
and assuming a conservative distance range of 15--50\,pc the space volume potentially occupied by the pair is 
$\lesssim1.22\times10^{-6}$~pc$^3$. Of all the {\it Gaia} DR2 stars within 100~pc and $\pm$20~deg of NLTT~51469A only 0.08\% 
share its proper motion at the 3$\sigma$ level, considering a 5~mas/yr error in the proper motion of the M3. Coupling this 
with the local space density of 0.06--0.11 stars/pc$^3$ \citep{2007AJ....133.2825R} we estimate that the probability of a 
chance alignment in space and motion for these two objects is less than $1.1\times10^{-10}$.
We can thus conclude that both stellar components are physically bound.

Using the absolute magnitude-spectral type and color-spectral type relations from \cite{2013ApJS..208....9P}, and the 
spectral type determination of the primary, we estimate that the secondary is an M6\,$\pm$\,1 dwarf from the derived relative 
magnitude differences in the four bands and the photometric colors: $I-J=2.31\pm0.33$, $I-K_s=3.04\pm0.32$, $J-K_s=0.73\pm0.05$.

\section{Distance}
The {\it Gaia} DR2 provided the parallax measurement for the primary NLTT 51469, $\pi$\,=\,21.457\,$\pm$\,0.611~mas, 
which translates to a distance of 46.6\,$\pm$\,1.3~pc. The object at 0\farcs64 from the primary was not resolved by {\it Gaia},
however, NLTT 51469 was tagged as a duplicated source in the DR2 catalog. The brown dwarf companion SDSS~2131--0119 was 
beyond the detection limit of {\it Gaia}. To assess whether it is located at a consistent distance we have estimated its 
spectrophotometric distance, assuming that the system has the age of the field. We used the VHS $J$, $H$, $K_s$ and 
{\it WISE W}1 and {\it W}2 photometry with near- and mid-infrared absolute magnitude versus spectral type relations for 
field L and T dwarfs defined in \cite{2012ApJS..201...19D} and \cite{2015ApJ...810..158F}. In Table~\ref{specphot_distance} 
we list the obtained estimates and their uncertainties, which take into account uncertainties in spectral type determination, 
intrinsic scatter of absolute magnitudes of a given spectral type and errors in photometry. The $\Delta$/$\sigma$ ratios 
quantify the difference between the parallactic distance of the primary and the estimated spectrophotometric distance of 
the L9 relative to the corresponding uncertainties. The values are consistent within the errors for all the considered 
bands and yield a mean distance of the L9 of 40\,$\pm$\,11~pc. This distance constraint for the L9 is consistent with 
the parallactic distance of NLTT~51469 at a level of 1$\sigma$.

We estimated the spectrophotometric distance also for the primary, for which we employed the DENIS $I$ and 2MASS $J, H, K_s$ 
photometry, since at these magnitudes the VHS starts to get beyond the linear regime of the detector. To account for 
the source at 0\farcs64 being unresolved in DENIS and 2MASS, we use the deblended magnitudes and 
by considering the mean absolute magnitudes for a given early-mid M sub-type from the compilation of \cite{2013ApJS..208....9P}
we find $d_I$\,=\,33\,$\pm$\,17~pc and $d_{JHK_s}$\,=\,$34^{+10}_{-13}$~pc. The large errors are mainly due to uncertainty 
in spectral type, since a range from M2 to M4 implies a $\sim$1.5~mag difference in brightness in these bands. The 
spectrophotometric distance of the primary is slightly lower but consistent with the parallactic distance to within the 2$\sigma$ 
uncertainty level. The spectrophotometric distance values of both objects also coincide, reinforcing that they are located at 
the same distance, which is consistent with companionship.

\begin{figure}
 \centering
 \includegraphics[scale=0.8,keepaspectratio=true]{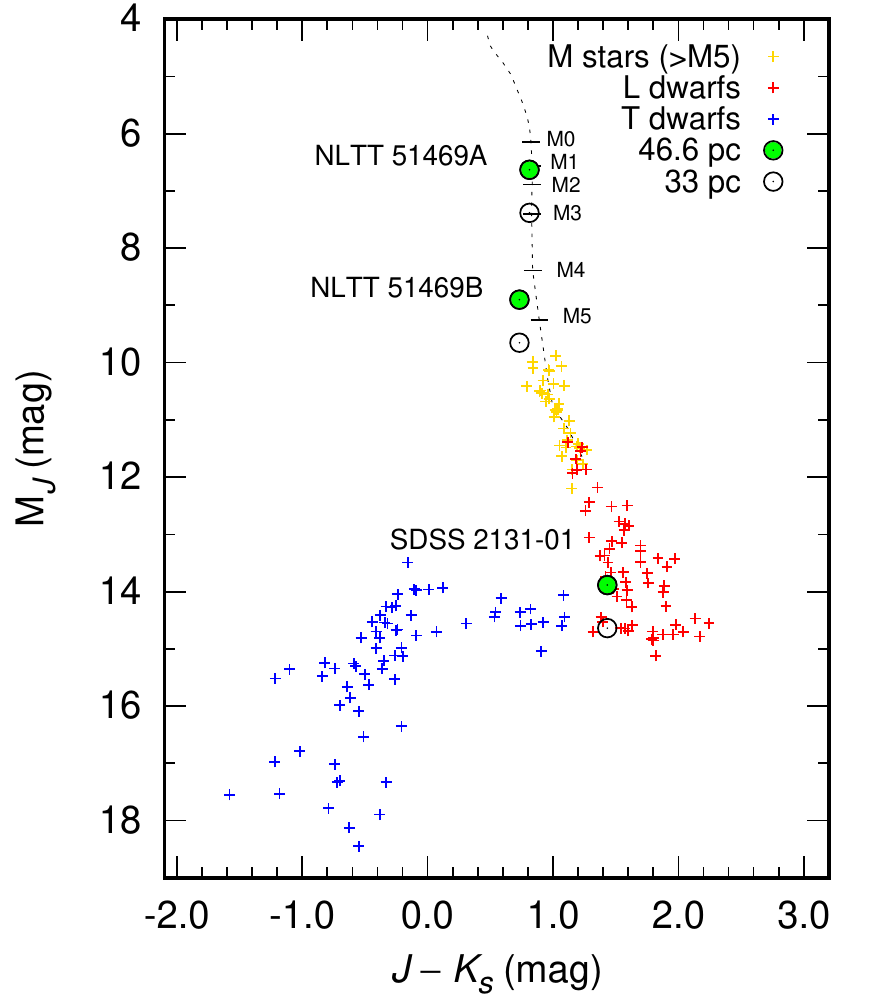}
 \caption{$M_J$ vs. $J-K$ color-magnitude diagram showing the two components of the NLTT~51469AB binary star 
 and its wide L9\,$\pm$\,1 brown dwarf companion overplotted against the sequence of K and M stars
 \citep{2013ApJS..208....9P} and ultracool dwarfs with measured parallactic distances
 from \protect\cite{2012ApJS..201...19D}. For the system we consider both the {\it Gaia} DR2
 distance of $d$\,=\,46.6\,$\pm$\,1.3~pc and our spectrophotometric estimation at $d\sim34$~pc.}
 \label{fig_color_mag}
\end{figure}

In our further analysis we consider both the {\it Gaia} distance to the system, i.e., 46.6\,$\pm$\,1.3~pc, and the 
spectrophotometric distance of $34^{+10}_{-13}$~pc. The corresponding projected orbital separations between the components are 
$\sim$2800--3800~au (M3 and L9) and 22--30~au (M3 and $\sim$M6). For these two distances, we compare the absolute $J$ magnitudes 
versus $J-K_s$ colors of the M3, $\sim$M6 and L9 components on the color-magnitude diagram in Figure~\ref{fig_color_mag} 
with a sequence of K and M stars \citep{2013ApJS..208....9P} and field late-M, L and T dwarfs with measured parallaxes compiled 
in \cite{2012ApJS..201...19D}.

The components follow the sequence and have photometric colors in agreement with those expected for their spectral 
type, but, adopting the 46.6~pc distance all the three objects appear about 1~mag brighter than their standard counterparts 
of the corresponding spectral type. The primary matches better to an M1--M2 and the secondary to an M4--M5 type dwarf. 
We suggest that this may be due to parallax determination being affected by the companion at 0\farcs64 unresolved 
by {\it Gaia}.

Evidence has been reported in other similar cases (e.g. 2MASS~J0249--0557AB, \citealt{2018AJ....156...57D}) that the 
five parameter {\it Gaia} DR2 astrometric solutions can be altered, in a systematic way, by the orbital motion of 
unresolved binaries. Particularly when a relatively small number of independent observation epoch was available (DR2 
reports that only 8 visibility periods were used in case of NLTT~51469). The astrometric excess noise of the source in 
DR2 is $\epsilon_i=2.74$\,mas at a significance of 2202$\sigma$, indicating that indeed the {\it Gaia} astrometry of 
this binary is most likely affected by correlated noise from orbital motion. The parallax is 15$\sigma$ lower than it 
would be expected from the spectrophotometric distance, but, as noted by \citealt{2018AJ....156...57D}, DR2 parallax 
systematics for unresolved binaries can be up to $\approx$20$\sigma$. The forthcoming Gaia data release with improved 
astrometry of non-single stars and with the binary information reported in our paper shall provide a more accurate
distance for the system.

\section{Companionship of NLTT~51469AB and SDSS~2131--01}
To evaluate whether the NLTT~51469 and SDSS~2131--01 objects are physically bound or if they have been found together at this 
separation due to chance alignment we estimated the probability of finding an L, T-type dwarf in our common proper motion search for 
companions and that both objects have consistent proper motion within 50~mas/yr.

To calculate the contamination rate of L, T dwarfs we need to determine the density of such objects in the VHS survey down to the 
limiting magnitude of 2MASS. To do this we have found that about 400 L and Ts are identified in the full VHS area ($\sim$20000~deg$^2$),
implying a surface density of 0.02 objects per square degree, and hence a probability for the presence of an L or T dwarf up 
to 20 arcmin around a star is 0.67\%. To estimate the probability that the companion is not physically related but has a 
consistent proper motion with the primary we have identified that 548 objects out of 50382 in our HPM catalogue have common 
proper motion with NLTT 51469, which represents a probability of 1.1\%.

Assuming a poissonian distribution, the probability of finding an L or T dwarf whose proper motion is consistent with a nearby
NLTT star in our VHS search is given by:
\begin{equation}
 P_{(x>0)}=1-P_{(x=0)}=1-e^{-\lambda}, 
  ~{\rm where}~~ \lambda = np
\end{equation}
where $P$ is the poissonian probability distribution, $n$ the number of stars and $p$ the combined probability of finding LT dwarfs with 
consistent proper motion. For our search we estimate that this probability is 26\% (it is relatively low, but not negligible). Considering 
that SDSS~2131--01 was found at angular separation of 82.3 arcsec, the probability to find such objects within this separation is much 
lower (0.14\%). In this calculation we have not taken into account that additionally, both objects appear to be located at a consistent 
distance. In summary we conclude that it is highly unlikely that the two components are unrelated objects found by chance alignment, and 
therefore that the NLTT~51469AB and SDSS~2131--01 is a physically bound multiple system.

\begin{figure}
 \includegraphics[width=\columnwidth,keepaspectratio=true]{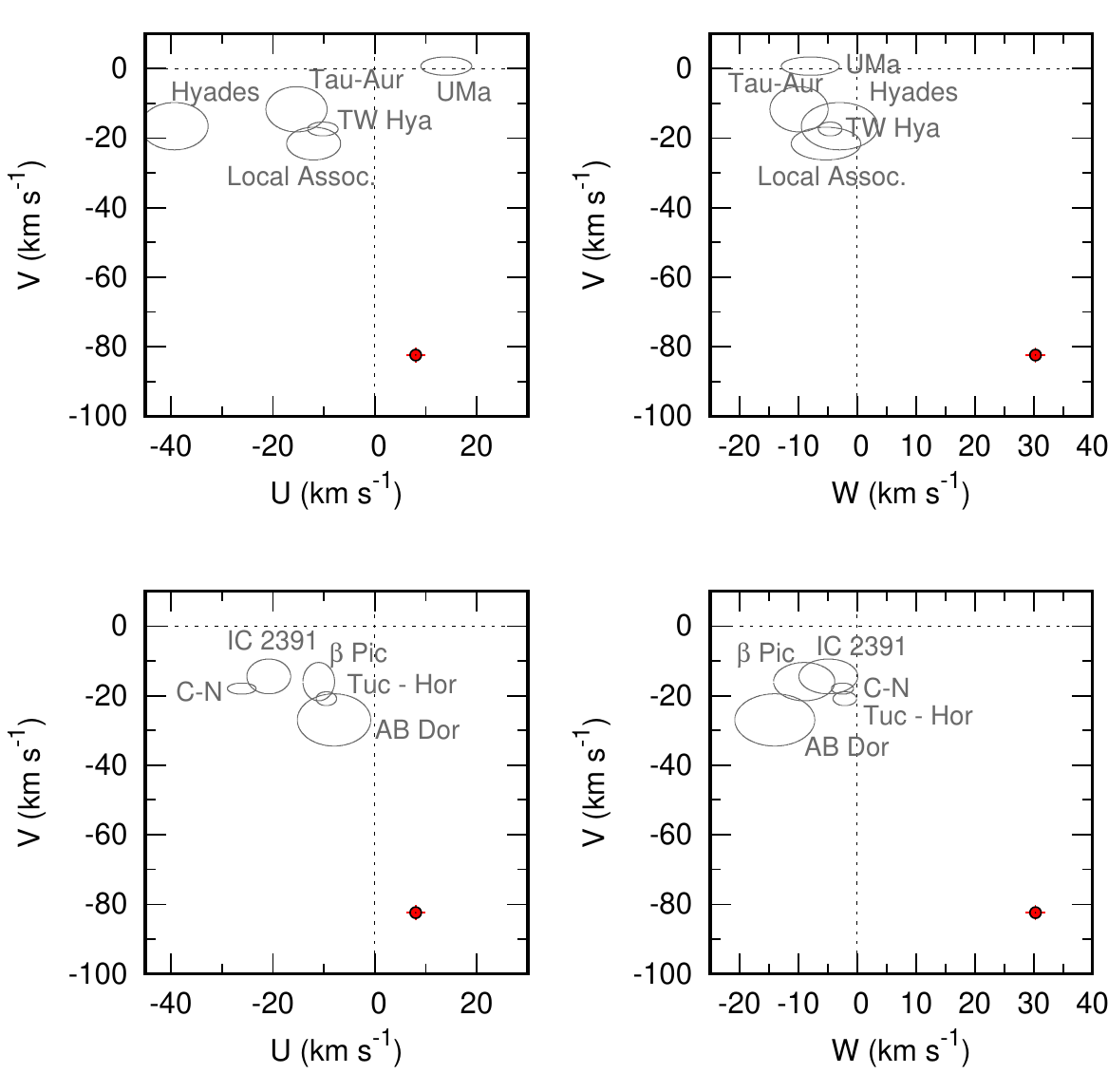}
  \caption{Galactic space velocities of NLTT 51469 (red dots) with overplotted 
  ellipsoids of known young star associations and moving groups. 
  Errors incorporate uncertainties in the proper motion, parallactic distance, and 
  radial velocity. Galactocentric {\it U} velocity is positive toward the Galactic center.}
  \label{uvw}
\end{figure}

\section{Physical properties}

\subsection{Radial velocity, galactic kinematics and age}
\begin{figure}
 \includegraphics[width=\columnwidth,keepaspectratio=true]{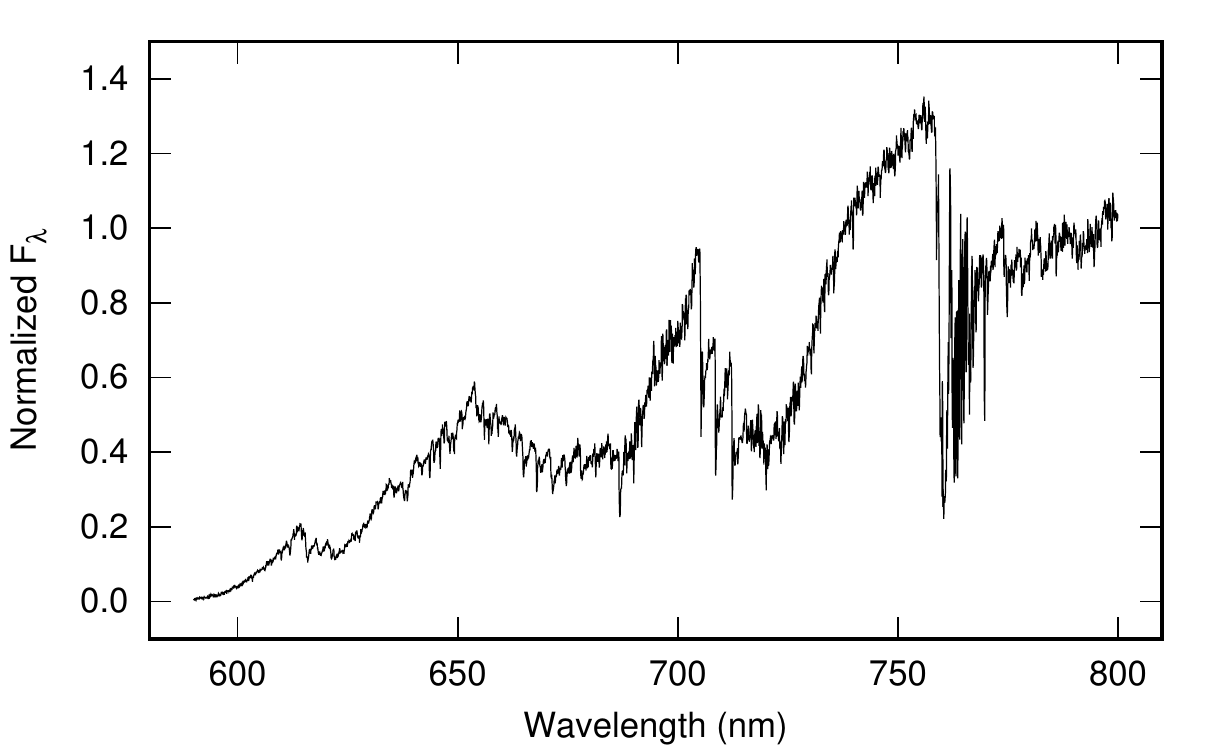}
 \hspace*{6mm}\includegraphics[width=0.9\columnwidth,keepaspectratio=true]{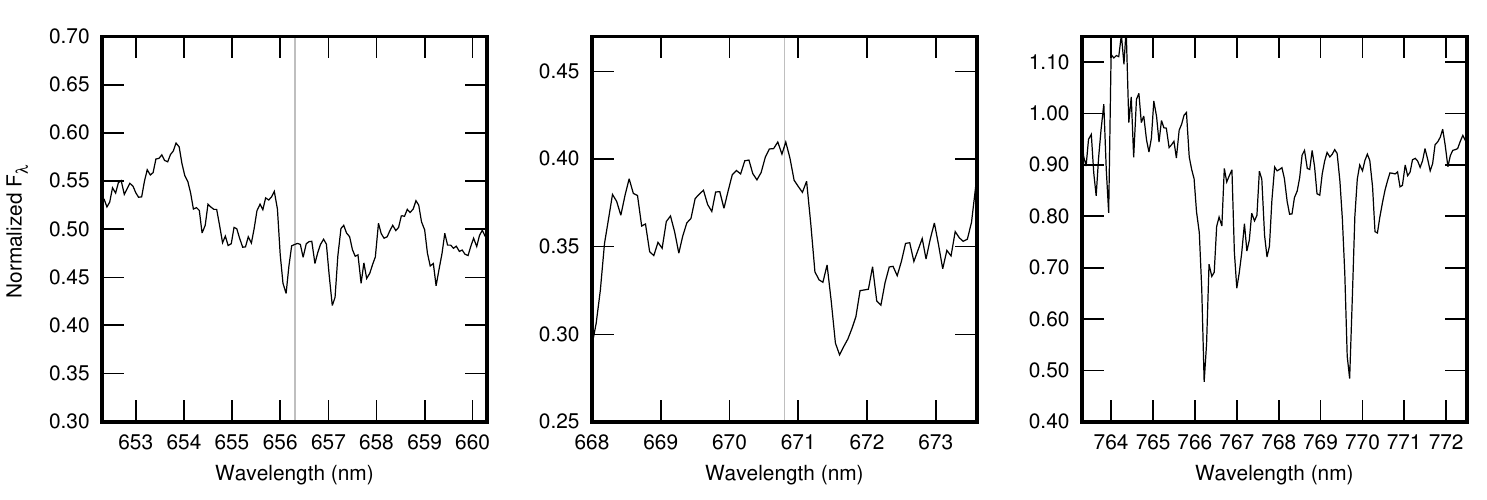}
 \caption{LT/FRODOSpec spectrum of NLTT~51469A covering 580--900\,nm wavelength range at a resolution of $R$\,$\sim$\,5300. 
  No telluric correction or flux calibration was applied to the displayed spectrum. Bottom plots show a close-up of $\lambda$ regions
  of H$\alpha$, Li{\sc I} and K{\sc I} doublet lines.}
 \label{FrodoSpec_red}
\end{figure}

We employed the LT/FRODOSpec intermediate-resolution red optical spectrum ($R$\,$\sim$\,5000, 600--800\,nm) of NLTT~51469A, displayed 
in Fig.~\ref{FrodoSpec_red}, to measure its heliocentric radial velocity, at the mean Modified Julian Date, MJD\,=\,57593.061387. We used 
the cross-correlation method against the M4.5V star GJ~876, which has a known, constant radial velocity of $v_h$\,=\,0.413\,$\pm$\,0.124 kms$^{-1}$ \citep{2002ApJS..141..503N}.
The cross-correlation was computed using the {\sc fxcor} task within {\sc iraf} over 6500--7500 and 7700--7950\AA ~wavelength range 
containing good signal-to-noise (S/N\,$>$\,50) data not affected by the telluric absorption. We fit a Gaussian function to the peak of 
the cross-correlation distribution. The resulting relative displacement was corrected for the lunar, diurnal, and annual velocities to 
obtain the heliocentric radial velocity of NLTT~51469A. We measured $v_h$\,=\,$-$64.3\,$\pm$\,9.0~kms$^{-1}$, 
where the error bar accounts for the uncertainties due to the cross-correlation procedure and the error associated with the velocity of 
the M4.5V standard star. 

The star has a previous radial velocity measurement obtained by \cite{2013AJ....146..134K} on October 17, 2009 (MJD\,=\,55121), 
$\sim$7~yr before our FRODOSpec observation. These authors obtained $v_h$\,=\,$-$67.34\,$\pm$\,2.76 kms$^{-1}$, which is consistent 
with our determination within the error bars. {\it Gaia} DR2 does not provide the radial velocity of the M3 primary, but it does provide 
$T_{\rm eff}$\,=\,3654~K. The $\sim$M6-type companion at 0\farcs64 induces a radial velocity variation of a maximum semi-amplitude of 
about 1~km/s over an orbital period of roughly 200~yr for an edge-on, circular orbit. This implies a shift of up to $\sim$10~m/s 
per year for an edge-on orbit, readily measurable with modern spectrographs. 

Having the proper motion, radial velocity and parallactic distance we calculated the three components of the Galactic 
space velocity, $U$, $V$, and $W$ of NLTT~51469A applying the formulas given by \cite{1987AJ.....93..864J}. We used the more precise 
literature value of $v_h$ for the calculation, and derived $U$, $V$, $W$\,=\,8.0\,$\pm$\,1.8, $-$82.4\,$\pm$\,2.1, 30.3\,$\pm$\,1.7~km\,s$^{-1}$.
The errors take into account the uncertainties of the proper motion, distance, and radial velocity. 
Figure~\ref{uvw} illustrates the ellipsoids corresponding to well-characterised young stellar moving groups of the solar neighborhood
(data compiled from \citealt{2004ARAA..42..685Z} and \citealt{2008hsf2.book..757T}) and the space velocity of NLTT 51469.
The high $V$ and $W$ indicate that the star does not belong to any of the nearby young moving groups, its velocities are compatible with 
velocity dispersions of the thin galactic disk population (\cite{1992ApJS...82..351L} and references therein).

\begin{table}
\centering
\caption{Measurements and determined parameters of the system\label{measurements}}
\scriptsize
\begin{tabular}{l c c}
\hline
\hline
Astrometry                                  & NLTT~51469                                             & SDSS~2131-0119 \\
\hline
R.A.  (J2000)                               & \hspace*{2pt}21$^{\rm h}$31$^{\rm m}$59$^{\rm s}$.603  & \hspace*{2pt}21$^{\rm h}$31$^{\rm m}$54$^{\rm s}$.391 \\
Decl. (J2000)                               & $-$01\degr20\arcmin06\farcs554                         &  $-$01\degr19\arcmin40\farcs511 \\
2MASS ID                                    & J213159.66-012003.9                                    & J213154.44-011937.4 \\
Separation (arcsec)$^a$                     & \multicolumn{2}{c}{82.27\,$\pm$\,0.02}   \\
Separation (AU)                             & \multicolumn{2}{c}{3834\,$\pm$\,100}     \\ 
Position angle (deg)$^a$                    & \multicolumn{2}{c}{288.4\,$\pm$\,0.1}    \\
$\mu_{\alpha}\cos\delta$ (mas\,yr$^{-1}$)   & $-$106\,$\pm$\,5 & $-$100\,$\pm$\,20     \\ 
$\mu_{\delta}$ (mas\,yr$^{-1}$)             & $-$235\,$\pm$\,5 & $-$230\,$\pm$\,20     \\
$\mu_{\alpha}\cos\delta$ (mas\,yr$^{-1}$)$^b$ & $-$95.49\,$\pm$\,0.96                    & ... \\ 
$\mu_{\delta}$ (mas\,yr$^{-1}$)$^b$           & $-$239.38\,$\pm$\,0.96                   & ... \\
Parallax $\pi$ (mas)$^b$                  & 21.45\,$\pm$\,0.61                       & ... \\
Estimated $d$ (pc)                        & $34^{+10}_{-13}$                                & $40\pm11$ \\
Parallactic $d$ (pc)                      & $46.6\pm1.3$                             & ... \\
$v_r$ (km\,s$^{-1}$)$^c$                  & $-67.34\pm2.76$                          & ... \\
                                          & $-64.3\pm9.0$                            & ... \\
{\it U} (km\,s$^{-1}$)                    & \hspace*{1.4mm}7.12\,$\pm$\,1.61         & ... \\  
{\it V} (km\,s$^{-1}$)                    & $-$82.56\,$\pm$\,2.12                    &  ...\\  
{\it W} (km\,s$^{-1}$)                    & \hspace*{1.4mm}30.16\,$\pm$\,1.80        & ... \\  
\hline
Photometry (mag)    & & \\
\hline
$V$            & $13.58\pm0.01$ & ... \\
$R$            & $13.47\pm0.09$ & ... \\
$I$            & $11.20\pm0.02$ & ... \\ 
$G$            & $12.677\pm0.001$ & ... \\
Sloan $g$      & $15.604\pm0.006$ & ... \\   
Sloan $r$      & $13.369\pm0.001$ & ... \\
Sloan $i$      & $12.040\pm0.001$ & $>$22.68 \\
Sloan $z$      & $13.116\pm0.013$ & $19.890\pm0.100$ \\
2MASS $J$      &  $9.848\pm0.029$ & $17.396\pm0.263$ \\
2MASS $H$      &  $9.281\pm0.021$ & $15.781\pm0.194$ \\
2MASS $K_s$    &  $9.045\pm0.023$ & $15.559\pm0.210$ \\
VHS $J$        &  $<$10.88        & $17.230\pm0.015$ \\
VHS $H$        &  $<$11.10        & $16.419\pm0.016$ \\
VHS $K_s$      &  $<$10.10        & $15.796\pm0.018$ \\
{\it WISE W}1  & $8.894\pm0.022$  & $15.075\pm0.037$ \\
{\it WISE W}2  & $8.736\pm0.019$  & $14.824\pm0.072$ \\
{\it WISE W}3  & $8.610\pm0.028$  & ... \\
{\it WISE W}4  & $8.123\pm0.294$  & ... \\
\hline
Spectral Classification  & & \\
\hline
Optical                & M3.5\,$\pm$\,0.5 & L9.0\,$\pm$\,1.0 \\ 
Near-IR                & M3.0\,$\pm$\,1.0 & \hspace*{1.9mm}L9.0\,$\pm$\,0.5\,$^d$  \\
Adopted spectral type  & M3.0\,$\pm$\,1.0 & L9.0\,$\pm$\,1.0 \\
\hline
Physical Properties & & \\
\hline
Age (Gyr)                     & \multicolumn{2}{c}{1\,--\,10} \\
$\log(L_{\rm bol}/L_{\odot})$ & $-1.50^{+0.02}_{-0.04}$  & $-4.4\pm0.1$ \\
$\log(L_{\rm bol}/L_{\odot})^e$ & $-1.78^{+0.02}_{-0.04}$  & $-4.7^{+0.3}_{-0.5}$ \\
$T_{\rm eff}$ (K)             & $3410^{+140}_{-210}$  & 1400--1650   \\
Mass ($M_{\odot}$)            & $0.42\pm0.02$         & 0.05--0.07   \\
\hline
\end{tabular}
\begin{flushleft}
{$^a$~Measured using VHS images, epoch (MJD) = 55373.363111} \\
{$^b$~From the {\it Gaia} DR2} \\
{$^c$~Literature value and our measurement} \\
{$^d$~Near-IR spectral type from \cite{2006AJ....131.2722C}} \\
{$^e$~Considering spectrophotometric distance}
\end{flushleft}
\end{table}

We used the FRODOSpec spectrum also to look for spectral features recognized as indicators of youth in very low-mass stars. We 
investigate the lithium content (Li~{\sc I} line at 670.8 nm), the H$_{\alpha}$ emission line at 656.3 nm and the potassium doublet 
at 766.5 and 769.9 nm. We did not detect the Li~{\sc I} absorption line with detection upper limit on the pseudo equivalent width 
(pEW) of 60 m\AA~ nor H$_{\alpha}$ emission at the level higher than $-$0.5~\AA. For the K~{\sc I} lines we measure pEWs of 
1.0\,$\pm$\,0.2 and 1.1\,$\pm$\,0.1~\AA. 

No significant H$_{\alpha}$ emission suggests low chromospheric activity, indicative of a spun down M dwarf. The absence of 
lithium and the relatively high V and W galactic velocities rule out youth. Also, the non-detection in X-rays by the ROSAT All 
Sky Survey and the GALEX UV detection with NUV\,=\,23.05\,$\pm$\,0.30~mag (2.2\,$\pm$\,0.6~$\mu$Jy) and non-detection in FUV 
indicate no significant flux excess, as compared to early M dwarfs at young ages until a few hundred Myr, and agrees with 
ages older than the Hyades at 650~Myr, for which a decay in UV and X-ray flux excess reaches a factor of 20 and 65 
\citep{2013MNRAS.431.2063S, 2014AJ....148...64S}. Together these criteria point to the NLTT~51469 system belonging to the thin 
disk population, with a likely age in the range of $\sim$1--5\,Gyr.

\subsection{Luminosity, mass and effective temperature}

We determined the bolometric magnitude and luminosity of the primary using $JHK_s$-band photometry from 2MASS, decomposed 
to account for an additional, unresolved $\sim$M6 source, and the bolometric corrections (BC) for field M dwarfs from 
Table~5 in \cite{2013ApJS..208....9P}. We employed the solar bolometric magnitude of 4.74~mag and obtained
$M_{\rm bol}$\,=\,$8.48^{+0.10}_{-0.05}$~mag and a luminosity of $\log(L_{\rm bol}/L_{\odot})$\,=\,$-1.50^{+0.02}_{-0.04}$~dex.
Errors in this determination include error in distance (parallax), spectral type and photometry. These values correspond to a rather 
earlier type at around M1.5, whereas for an M3$\pm$1 dwarf one would expect $M_{\rm bol}$\,$\sim$\,9.2$^{+1.0}_{-0.5}$~mag 
and a luminosity of $-1.78^{+0.2}_{-0.4}$~dex. As mentioned earlier, this discrepancy may be a result of an error 
in the parallax determination due to the stellar companion at 0.6 arcsec.

For the estimation of effective temperature and mass we also used the values from the compilation of \cite{2013ApJS..208....9P}, 
which for an M3\,$\pm$\,1 dwarf yield a $T_{\rm eff}$\,=\,3410$^{+140}_{-210}$~K and $m$\,=\,0.36$^{+0.08}_{-0.14}$~$M_{\odot}$.
The {\it Gaia} parallactic distance implies a higher luminosity than expected for an M3 dwarf and hence mass estimate based 
on the mass-luminosity relation for main sequence M dwarfs \citep{2016AJ....152..141B} yields a consequently higher range of 
$m$\,=\,0.42\,$\pm$\,0.02~$M_{\odot}$. As for the $\sim$M6 component, based on the values from \cite{2013ApJS..208....9P} 
as reference values for a given spectral type we estimated $T_{\rm eff}$ and mass of 2850\,$\pm$\,200~K and 
0.10$^{+0.06}_{-0.01}$~$M_{\odot}$.

For the L9 companion we determined the bolometric magnitude and luminosity using the BC for field-age ultracool dwarfs
from \cite{2015ApJ...810..158F} and $J$-band photometry from VHS. Considering $d$\,=\,46.6\,$\pm$\,1.3~pc we obtained 
$M_{\rm bol}$\,=\,15.71\,$\pm$\,0.25~mag and $\log(L_{\rm bol}/L_{\odot})=-4.4\pm0.1$~dex, whereas for 
$d$\,=\,$34^{+10}_{-13}$~pc we got 16.40$^{+1.17}_{-0.69}$~mag and -4.66$^{+0.28}_{-0.47}$~dex, taking into account 
errors in distance, spectral type, photometry and intrinsic scatter of absolute magnitudes at a given spectral type. 
We then employed the AMES Dusty evolutionary model isochrones \citep{2001ApJ...556..357A, 2011ASPC..448...91A, 
2012RSPTA.370.2765A} to infer the $T_{\rm eff}$ and mass of the companion for three different ages and solar abundance, 
using the {\it Gaia} distance. We obtained masses between 0.050--0.055, 0.068--0.071 and 0.068--0.071~M$_{\odot}$ 
and temperatures in the range of 1400--1550, 1450--1600 and 1500--1650~K for 1, 5, and 10~Gyr, respectively. From the 
polynomial relations of $L_{\rm bol}$ and $T_{\rm eff}$ as a function of spectral type for field age objects determined 
by \cite{2015ApJ...810..158F} the expected luminosity and temperature of an L9-type object is $-$4.55\,$\pm$\,0.13~dex 
and 1300\,$\pm$\,120~K, respectively. These values are somewhat lower than the ones derived using bolometric corrections, 
$J$-band photometry and parallactic distance, but consistent within the errors.

\section{Conclusions and future prospects}
Using the VHS and 2MASS surveys and follow-up imaging and spectroscopic observations we have identified a very low-mass 
hierarchical triple system, NLTT~51469AB/SDSS~2131--0119, composed of a close stellar binary at an angular separation 
of 0.64\,$\pm$\,0.01 arcsec (projected distance of $\approx$\,30\,au), and a wide (82.27\,$\pm$\,0.02 arcsec, 
$\approx$\,3800\,au) co-moving brown dwarf companion. We determined the spectral type of the primary NLTT~51469A and 
confirm the spectral type of the brown dwarf as an M3$\pm$1 and L9$\pm$1, respectively. We have also estimated 
the spectral type of the close companion NLTT~51469B to be $\sim$M6$\pm$1. 

The {\it Gaia} measurement of the parallax of NLTT~51469 yields a distance of 46.6\,$\pm$\,1.3~pc. Our spectrophotometric 
distance estimates are compatible with this, though indicate the possibility of a slightly closer distance (that may be 
consistent with {\it Gaia} uncertainty induced by multiplicity). The matching proper motions and agreement in distance of 
the M3 and L9 components, and the compatible angular separations and position angles of the $\sim$M6 component measured 
at three epochs lead us to the conclusion that the three objects form a physically bound system.

We determined the luminosities, effective temperatures and masses of the three components. It is worth noting that an age 
near 1~Gyr (at the younger end of the thin disk range) would imply a mass of the L9 brown dwarf at or below 0.055~M$_{\odot}$, 
and thus preservation of lithium \citep{1993ApJ...404L..17M}. Intermediate resolution spectroscopy covering the Li\,{\sc i} 
line at 670.8~nm could thus be useful in providing additional age constraints.

Assuming masses of 0.4~$M_{\odot}$ for the M3, 0.1~$M_{\odot}$ for the M6, and 0.065~$M_{\odot}$ for the L9, and given 
the $\sim$3800\,au separation of the L9 companion this is certainly one of the systems with the lowest gravitational 
binding energy ($E_b$\,$\approx$\,$-$2.48\,$\times$\,10$^{42}$ erg). Yet, it can be energetically stable according to 
Fig.~16 of \cite{2007ApJ...660.1492C}. This might be telling us that the system was not formed in a dense environment, 
otherwise, encounters with other stars would have disrupted the least massive component. Another possibility is that the 
system we see today is the result of capture(s) from gravitational interactions with nearby or passing low mass sources. 
Only by characterizing the long- and short-period orbits and performing an exhaustive analysis of the chemical composition 
of each individual member of the system, we might be able to assess their origin. The presented system builds up the sample 
of benchmark objects for studies of the least massive stars and substellar objects, in particular, of brown dwarfs at the 
L/T transition.

\section*{Acknowledgements}
We sincerely thank the reviewer for his insightful comments that allowed to greatly improve the manuscript.

B.G. acknowledges support from the CONICYT through FONDECYT Postdoctoral Fellowship grant N$^{\rm o}$ 3170513.
This work is partly financed by the Spanish Ministry of Economy and Competitivity through the projects AYA2016-79425-C3-2-P.
N.L. and V.J.S.B acknowledge support from the Spanish Ministry of Economy and Competitivity through the project
AYA2015-69350-C3-2-P. A.P. acknowledges support from the Spanish Ministry of Economy and Competitivity through the project
AYA2015-69350-C3-3-P.
Based on observations obtained as part of the VISTA Hemisphere Survey, ESO programme, 179.A-2010 (PI: McMahon)
Based on observations collected at the European Organisation for Astronomical Research in the Southern Hemisphere
under ESO programme 092.C-0874(B).
Based on observations made with the Nordic Optical Telescope, operated by the Nordic Optical Telescope Scientific Association 
at the Observatorio del Roque de los Muchachos, La Palma, Spain, of the Instituto de Astrof\'isica de Canarias.
This paper includes data obtained using the 6.5\,m Magellan Clay Telescope at Las Campanas Observatory, Chile.
This publication makes use of data products from the Two Micron All Sky Survey, which is a joint project of the University of 
Massachusetts and the Infrared Processing and Analysis Center/California Institute of Technology, funded by the National 
Aeronautics and Space Administration and the National Science Foundation.
This publication makes use of data products from the Wide-field Infrared Survey Explorer, which is a joint project of the
University of California, Los Angeles, and the Jet Propulsion Laboratory/California Institute of Technology, funded by the 
National Aeronautics and Space Administration.
This work has made use of data from the European Space Agency (ESA) mission {\it Gaia} (\url{https://www.cosmos.esa.int/gaia}), 
processed by the {\it Gaia} Data Processing and Analysis Consortium (DPAC, \url{https://www.cosmos.esa.int/web/gaia/dpac/consortium}). 
Funding for the DPAC has been provided by national institutions, in particular the institutions participating in the {\it Gaia} 
Multilateral Agreement.
This research has made use of NASA's Astrophysics Data System.
We have made use of the ROSAT Data Archive of the Max-Planck-Institut f\"ur extraterrestrische Physik (MPE) at Garching, Germany.
This research has made use of the Washington Double Star Catalog maintained at the U.S. Naval Observatory.




\bibliographystyle{mnras}
\bibliography{mnras_manuscript} 

\bsp	
\label{lastpage}
\end{document}